\documentclass[a4paper,11pt]{article}

\usepackage{pos}
\usepackage{lipsum} 
\usepackage{titlesec}
\titlespacing*{\section}{0pt}{1ex plus 0.5ex minus 0.2ex}{0.5ex}

\usepackage{caption} 
\usepackage{natbib}
\usepackage{etoolbox}
\patchcmd{\thebibliography}{\settowidth}
  {\setlength{\itemsep}{0pt plus 0.1ex}
   \setlength{\parsep}{0pt}
   \settowidth}{}{}
   
\title{Status of the GRAND project}

\ShortTitle{GRAND status}

\author*[a, b, c]{Olivier Martineau-Huynh}
\onbehalf{for the GRAND Collaboration{\normalsize \\ \normalfont(a complete list of authors can be found at the end of the proceedings)}\\}

\affiliation[a]{Sorbonne Université, Université Paris Diderot, Sorbonne Paris Cité, CNRS, Laboratoire de Physique Nucléaire et de Hautes Energies (LPNHE), 4 place Jussieu, 75005 Paris, France}
\affiliation[b]{Institut d'Astrophysique de Paris, CNRS, Sorbonne Université, 98 bis bd Arago, 75014 Paris, France}
\affiliation[c]{National Astronomical Observatories, Chinese Academy of Sciences, Beijing 100101, China}

\emailAdd{omartino@in2p3.fr}

\abstract{

GRAND (the Giant Radio Array for Neutrino Detection) is a proposed next-generation observatory targetting primarily the detection of ultra-high-energy neutrinos, with energies exceeding about 100 PeV. GRAND is envisioned as a collection of large-scale ground arrays of self-triggered radio antennas that target the radio emission from extensive air showers initiated by UHE particles. Three prototype arrays are presently in operation: GRANDProto300 in China, with 65 units running since end of 2024, GRAND@Auger in Argentina with 10 units deployed on the site of the Pierre Auger Observatory, and GRAND@Nançay in France, a 4-unit setup installed at the Nançay radio-observatory and used for test purposes. The main objective of the GRAND prototype phase is to validate the detection principle and technology of GRAND, in preparation for its next phase, GRAND10k. GRAND10k will consist of two arrays of 10'000 antennas each, covering both the Northern and Southern hemispheres, to be deployed from 2030 on. Here we give an overview of the GRAND concept, its science goals, the status of the prototypes, their performances and first detection of cosmic rays, and the technical perspective they open for the future. 

\vspace{4mm}

}

\ConferenceLogo{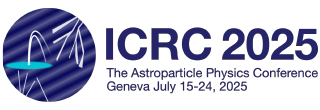}

\FullConference{39th International Cosmic Ray Conference (ICRC2025)\\  15--24 July, 2025\\ Geneva, Switzerland}

\begin{document}

\maketitle

\section{Introduction}\label{intro}
GRAND is a next-generation radio array designed to detect pimarily ultra-high-energy neutrinos via the radio emission of extensive air showers (EAS). With a staged deployment strategy, starting with prototypes GRANDProto300 (GP300) and GRAND@Auger (G@A), the project aims to validate its detection principle and prepare for large-scale deployment. This contribution presents an overview of the current status of GRAND and its scientific prospects: in Section\,\ref{science}, we outline the scientific objectives of GRAND. Section\,\ref{principle} presents the detection principle, while Section\,\ref{soft} discusses simulation and reconstruction developments. Section\,\ref{proto} reviews the status of the prototypes, and Section\,\ref{next} describes the next steps toward GRAND10k and the full GRAND array.

\section{The GRAND science case}\label{science}
Ultra-high-energy (UHE) neutrinos ($E \geq 10^{17}$\,eV) are expected to play a major role in time-domain multi-messenger astronomy. As undeflected signatures of hadronic acceleration processes~\cite{Ackermann:2022rqc} and thanks to their extremely low interaction cross-sections —which allow them to probe both the deep Universe and dense astrophysical sources— UHE neutrinos may in particular hold the key to solving the long-standing mystery of the origin of ultra-high-energy cosmic rays (UHECRs).

The recent detection of an UHE neutrino by KM3NeT~\cite{KM3NeT:2025npi} marks a major milestone. However, the expected flux at these energies is extremely low, requiring much larger detection areas than those currently in operation to obtain statistically significant event samples. The goal of GRAND is precisely to detect such UHE neutrinos~\cite{wp}, providing an  observational window distinct to IceCube and KM3NeT, whose sensitivities drop significantly above the PeV scale.

GRAND is designed to reach sensitivities matching the fluxes expected for cosmogenic neutrinos\,\cite{AlvesBatista:2018zui}. This deep sensitivity is achieved within a narrow field of view —a direct consequence of GRAND’s detection principle (see Section~\ref{principle})— which, combined with $\sim0.1^{\circ}$ angular resolution\,\cite{adf}, makes GRAND a highly effective instrument for the search for (transient) point sources~\cite{Kotera:2025jca}.

In addition, GRAND’s gigantic detection area —20 times that of the Pierre Auger Observatory in its baseline design~\cite{wp}— makes it a powerfull tool to study cosmic rays at the highest energies. GRAND also offers appealing perspectives for the study of neutrino properties, dark matter, Lorentz Invariance Violation\,\cite{wp} and tests of quantum gravity~\cite{2025CQGra..42c2001A}. The expected 200\,k antennas may also constitute a powerful instrument for fast radio burst (FRB) detection, with the added advantage that unphased signal summation enables a wide field of view, favorable to the serendipitous discovery of bright FRBs or radio giant pulses\,\cite{giantpulses}.
Finally GRAND will also enable the study of solar flares ---already observed in GP300\,\cite{gp300}---, lightning or thunderstorms~\cite{2003ligh.book.....R}.

\begin{figure}[t!]
\centering
\includegraphics[width=0.49\linewidth]{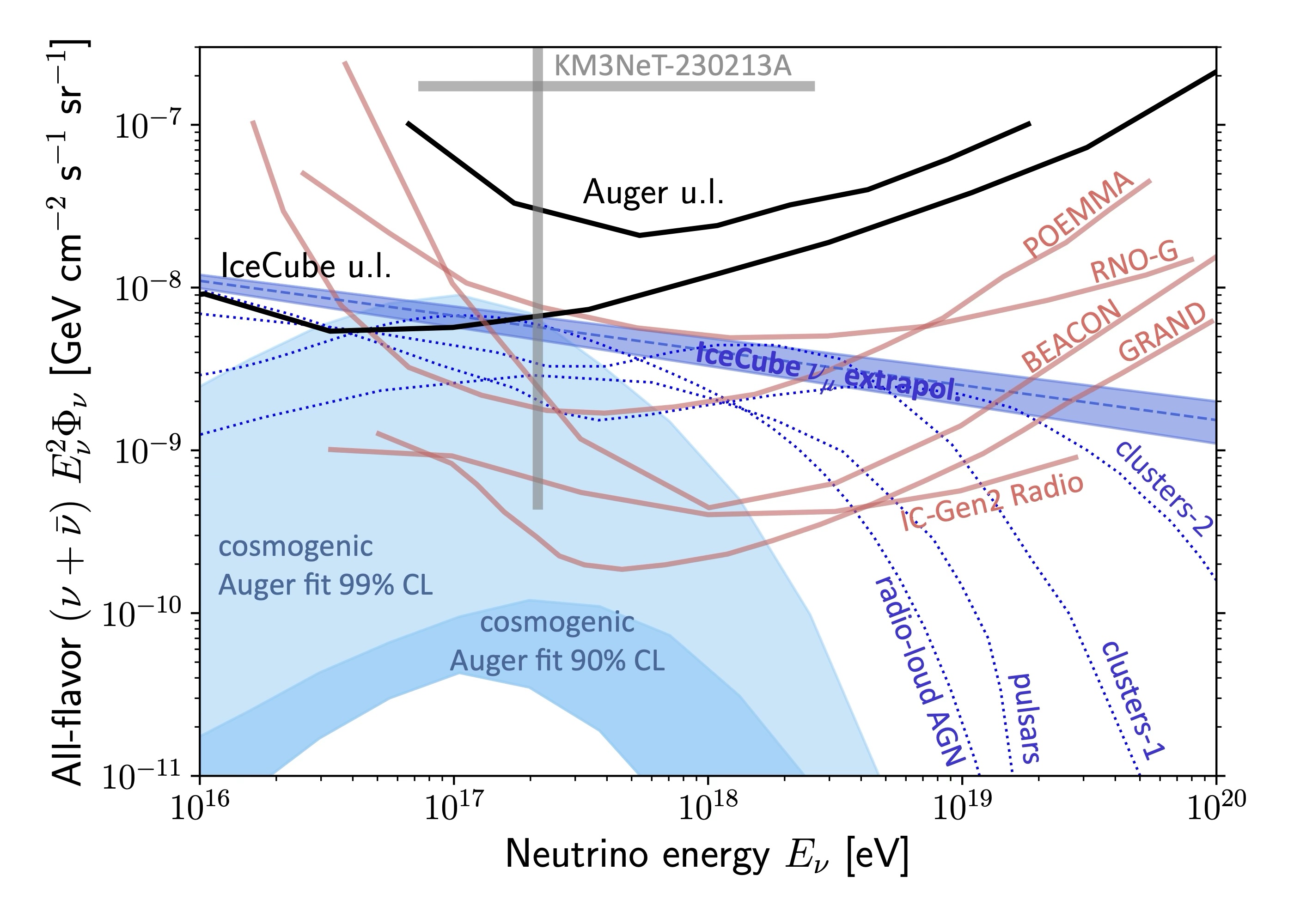}
\includegraphics[width=0.49\linewidth]{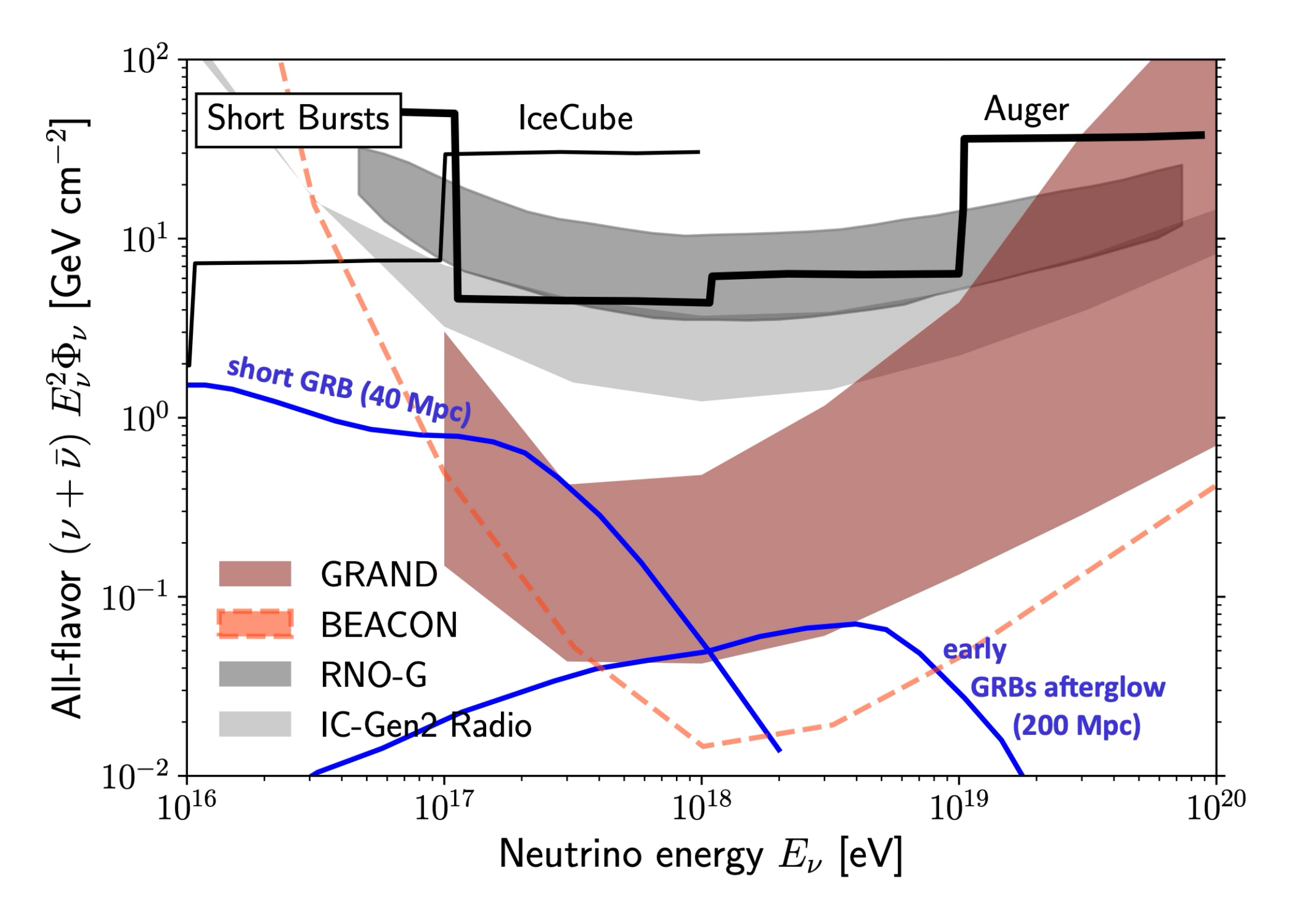}
\caption{Left: diffuse UHE neutrino fluxes
from various astrophysical (blue dotted lines) and cosmogenic origins. The pink solid lines indicate the projected 10-year differential sensitivities of GRAND and of several projects. Black solid lines mark the upper limits on UHE neutrinos from IceCube and Auger. 
Right: all-flavor neutrino fluence sensitivities per decade in energy for an assumed E$^{-2}$ neutrino spectrum for GRAND and other upcoming experiments. For Auger, IceCube, and IceCube-Gen2 the upper-limit sensitivity to neutrino transients at the location of GW170817A is shown. Overlaid are theoretical neutrino fluences for a short GRB/binary neutron star merger at 40\,Mpc and an early GRB afterglow at 200\,Mpc. Adapted from\,\cite{Kotera:2025jca}, where references for the curves displayed can also be found. }
\label{fig:sens}
\end{figure}

\section{The GRAND detection principle}
\label{principle}

Radio detection of extensive air showers (EAS) has matured into a reliable technique. It targets the coherent electromagnetic radiation emitted in the tens of MHz frequency range, caused by the deflection of electrons and positrons in the Earth's magnetic field —a phenomenon known as the geomagnetic effect. This effect is the dominant source of the short-duration ($\sim$100 ns), transient electromagnetic pulses detectable in this frequency range for showers with energies above $\sim 10^{16.5}$\,eV\,\cite{Huege:2016veh, Schroder:2016hrv}. Radio detection benefits from the valuable experience gained through previous experiments such as AERA, LOFAR, CODALEMA, Tunka-Rex, and TREND, which demonstrated that arrays of antennas can detect EAS in stand-alone mode\,\cite{trend}. 

GRAND's goal is to observe UHE neutrinos through radio detection. The core principle, detailed in\,\cite{wp}, is as follows: the neutrino interaction length inside Earth decreases from $\sim$6000 km at PeV energies to $\sim$100 km at EeV, effectively rendering the Earth opaque to them. When a neutrino undergoes a charged-current interaction, it produces a high-energy charged lepton of the same flavor. Tau leptons following Earth-skimming trajectories have a significant probability of emerging into the atmosphere and decaying quickly enough to generate a nearly-horizontal EAS. It will —just like any other EAS— emit radio pulses which can be detected by ground-based antennas.

The baseline design of the complete GRAND detector consists of a network of $\sim$20 arrays of 10,000 self-triggered radio antennas each, deployed in favorable locations worldwide\,\cite{wp}, totaling an ambitious 200,000\,km$^2$ detection area. The relatively low cost, ease of deployment, robustness, and stability of radio antennas opens the way for such a large-scale plan in the next decade. Nevertheless, ongoing intensive R\&D efforts (see e.g. \cite{heron}) aimed at improving detector sensitivity may enable the scientific goals outlined in Section \ref{science} to be achieved with a significantly smaller setup.

\section{GRAND simulations and reconstruction}
\label{soft}
In its early stage, the GRAND collaboration focused on developing a realistic end-to-end simulation based on custom tools, enabling the computation of the full GRAND detector's effective area (see \cite{wp}) and an assessment of its potential for neutrino detection (see e.g. Fig.\,\ref{fig:sens}).

Our efforts then shifted towards preparing the prototype stage of the GRAND project. In parallel with the instrumental developments described in Section \ref{proto}, considerable effort was devoted to simulating the response of the GRAND prototype detector to air showers. This includes a detailed description and calibration of the radio-frequency detection chain, enabling us to reproduce the detector’s response to stationary background noise with excellent precision (see Fig.\,\ref{fig:gal}).

These simulations, stored in a format identical to the experimental data, have enabled the development, testing, and refinement of reconstruction methods designed within the GRAND framework. These include the computation of the electric-field signal via deconvolution of the antenna response from recorded voltage traces\,\cite{kewen}, procedures to qualify and monitor the detector and its environment\,\cite{gp300}, algorithms to identify cosmic-ray candidates in the data\,\cite{cr_search}, and techniques to reconstruct their properties, achieving resolutions better than 0.1$^{\circ}$ in arrival direction and 15\% in energy\,\cite{ldf, adf, adf_v,gnn}. These developments are — or will be — integrated into \href{https://github.com/grand-mother/grand}{\sc GRANDlib}\,\cite{grandlib}, the open-source library of GRAND software tools.

\section{The GRAND prototype phase}
\label{proto}
The GRAND collaboration follows a staged approach to achieve its ambitious objectives. The current phase aims to demonstrate the feasibility of GRAND: efficient and pure radio detection of EAS down to nearly horizontal trajectories, along with a precise reconstruction of their properties (origin direction, energy, and nature). To this end, two prototypes are being used: GRANDProto300 and GRAND@Auger, described in\,\cite{gp300,gaa} while we provide below only a brief overview. Additionally, a four-unit setup deployed at the Nançay radio-observatory is used for test purposes\,\cite{Correa:2023Tn}. 

\subsection{Design of the GRAND prototypes}
\label{du_design}
\begin{figure}[t!]
\centering
\includegraphics[width=0.25\linewidth]{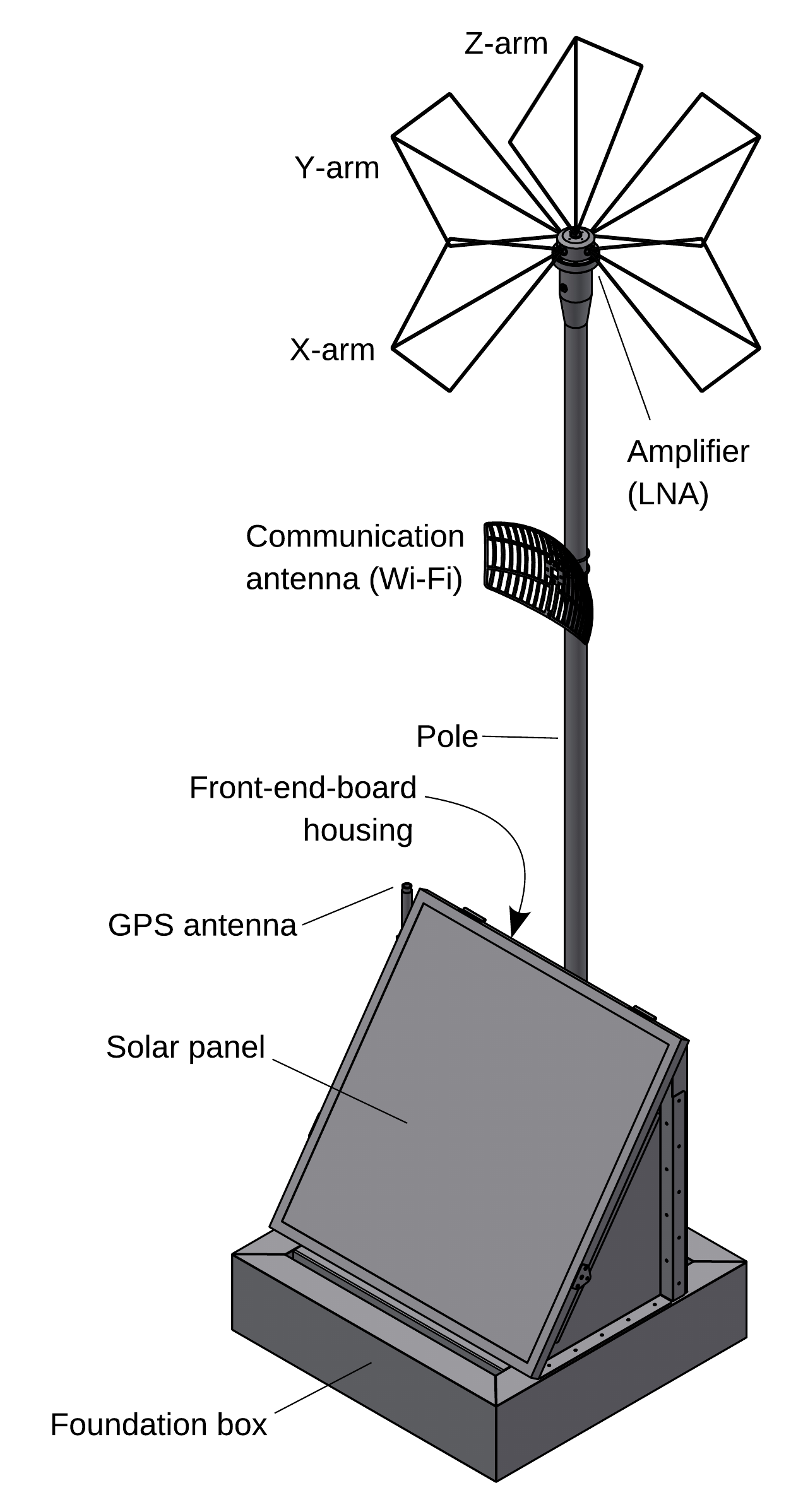}
\includegraphics[trim={0 0 0 0cm},clip,width=0.5\linewidth]{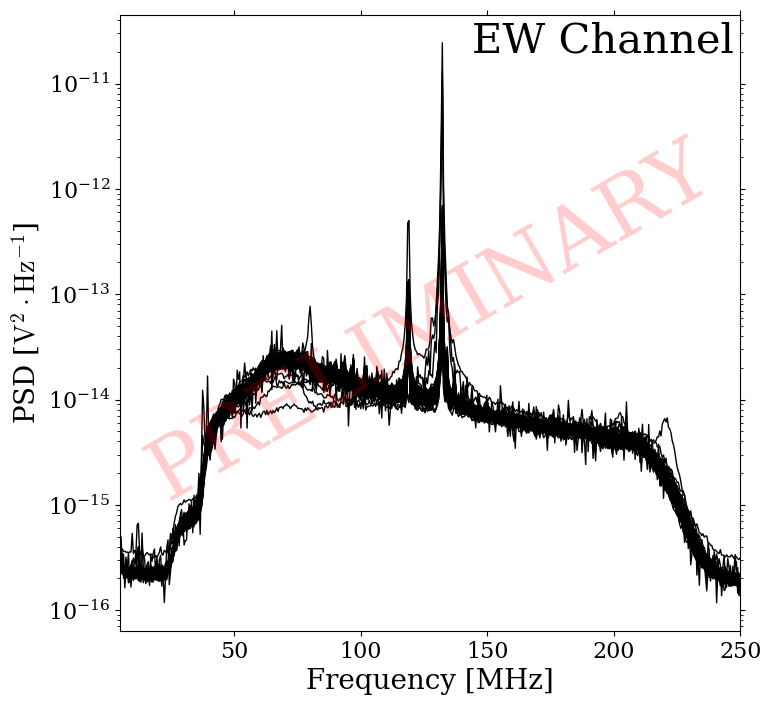}
\caption{Left: schematic diagram of a GP300 detection unit. The foundation box is filled with sand to weigh down the structure. Taken from \,\cite{grandlib}. Right: power spectrum density curves recorded on June 22, 2025 on the East-West arms of the 65 DUs deployed on the GP300 site The narrow lines between 119 and 136\,MHz correspond to aeronautic communications. Plot by Xishui Tian, Peking U. }\label{fig:DUfig}
\end{figure}

The detection units (DUs) of the GRAND prototypes share a very similar design. A schematic diagram of the model deployed on the GP300 site is presented in Fig.\,\ref{fig:DUfig}. Each unit features an antenna equipped with three independent radiators oriented along the East–West, North–South, and vertical directions. Their sensitivity is optimal in the 30–200\,MHz frequency range. The three signals from this so-called {\it Horizon Antenna} are amplified by low-noise amplifiers located directly within the antenna head, mounted at a height of 3.5\,m to reduce attenuation for signals propagating near the ground.

The signals are transmitted to a Front-End Board (FEB) located at the base of the antenna pole, powered by a battery recharged via a 150\,W solar panel. Antenna signals are filtered in the 50–200\,MHz range and further amplified to match the dynamic range of a 14-bit, 500\,MSamples/s ADC used for digitization. A Xilinx System-on-Chip, which includes an FPGA and four CPUs, then executes a first-level online trigger to select transient pulses after additional digital filtering (e.g., notch or FIR filters). Each trigger is timestamped using a Trimble GPS module integrated into the FEB and transmitted via Wi-Fi to a central DAQ system.

When multiple DUs —typically three to five— register coincident signals within a causal time window, a second-level trigger is issued by the central DAQ. In response, 2\,$\mu$s-long waveforms are recorded for the three channels of each triggered antenna. The raw data are subsequently transferred to the IN2P3 computing center and stored in a database accessible to GRAND members for offline processing. For more information on data formats and handling, see\,\cite{grandlib}.  

\subsection{GRANDProto300}
\begin{figure}[t!]
\centering
\includegraphics[width=0.48\linewidth]{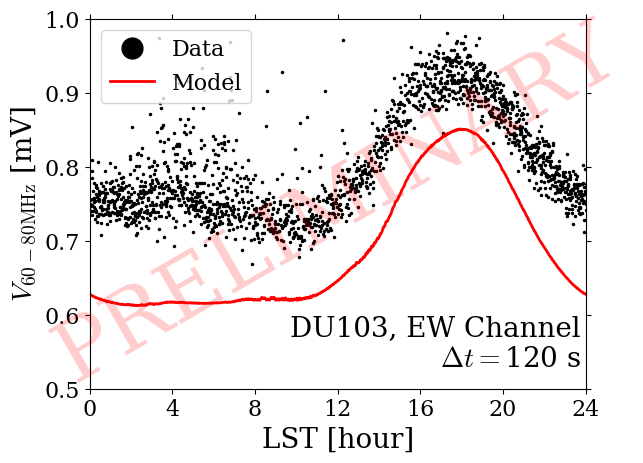}
\includegraphics[width=0.48\linewidth]{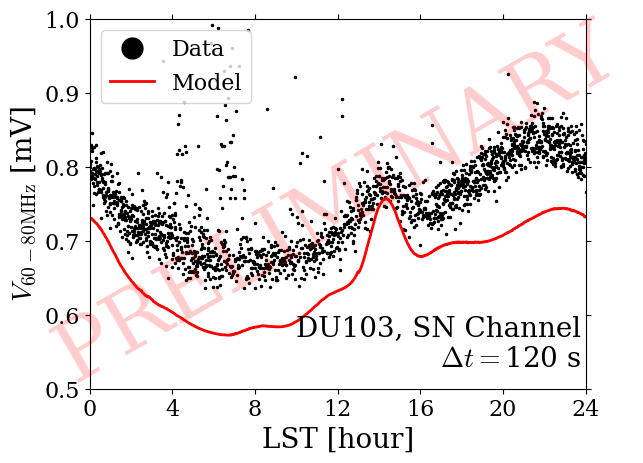}
\caption{Standard deviation of the stationary noise measured on one GP300 DU on June 22-25, 2025 in the frequency range 60-80\,MHz for East-West (left) and North-South (right) channels as a function of Local Sidereal Time. Each point corresponds to 24\,$\mu s$ of minbias data recorded over a period of 120\,s. A fraction smaller than 1\% of noisy data is excluded from these plots. The red curve is the expected contribution from the sky background only, computed at ADC level from the LFMap emission model\,\cite{lfmap} and simulation of the antenna + RF chain response, without any adjustment. This allows to estimate the electronic noise to $\sim$15-20\% for X and Y channels. Plot by Xishui Tian, Peking U. and Xin Xu, Xidian U.}\label{fig:gal}
\end{figure}

\begin{figure}[t!]
\centering
\includegraphics[width=0.37\linewidth]{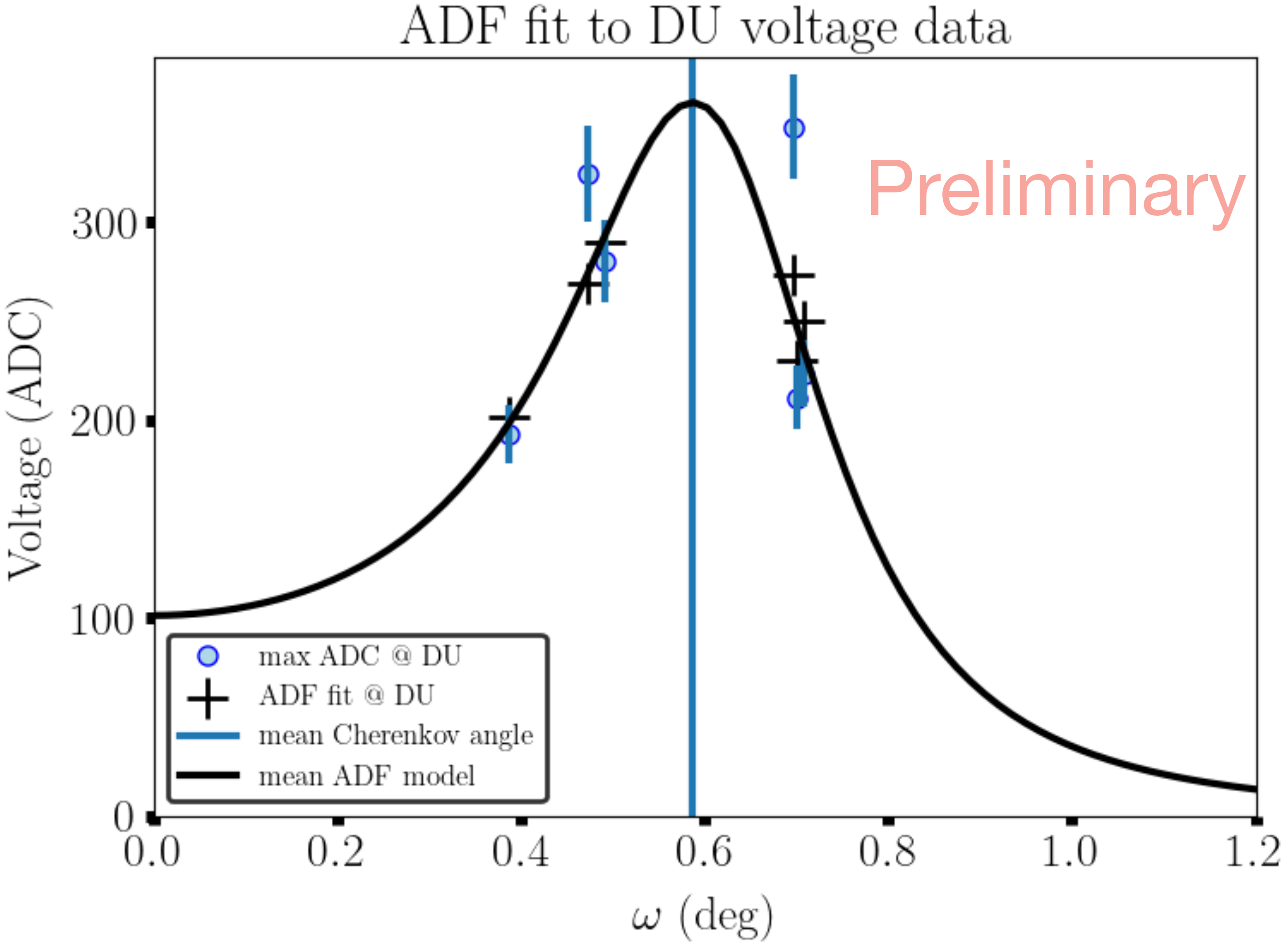}
\includegraphics[trim={0 0 0cm 4.5cm},clip,width=0.52\linewidth]{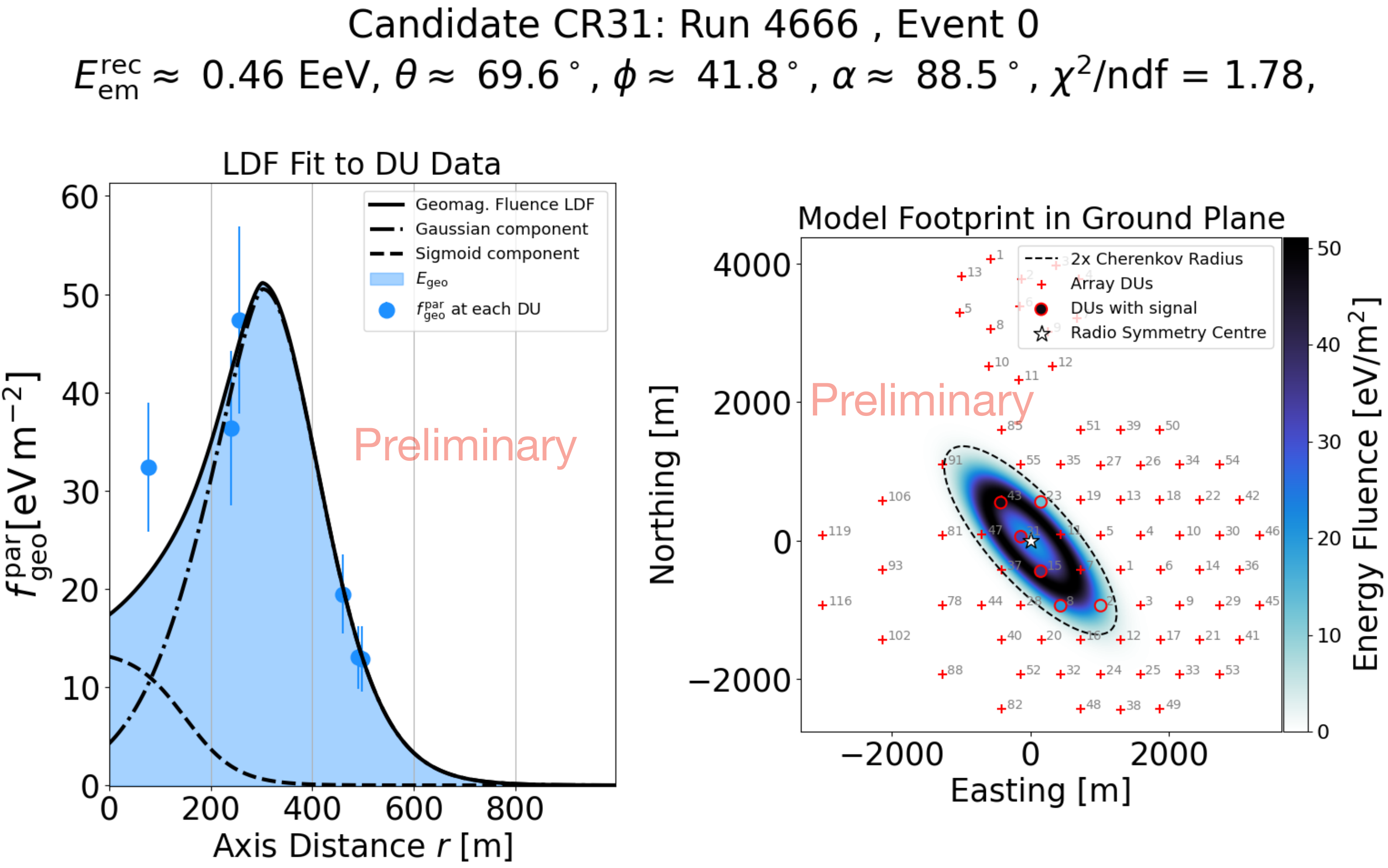}
\caption{A cosmic-ray candidate detected with GP300 on January 2, 2025. The left panel displays the Angular Distribution Function\,\cite{adf_v} fit to voltage as a function of angular distance to the shower axis. The central panel shows the fit from the Lateral Distribution Function\,\cite{ldf} to the energy fluence calculated from the estimated electric field\,\cite{kewen}. The event footprint is shown in the right panel, with a color map coding the amplitude distribution corresponding to the LDF fit. Triggered antennas are marked with a circle. Note that not all GP300 DUs (marked as red crosses in the plot) were running at the time of the event. 
Plots by M. Guelfand, Sorbonne U., and L. Gülzow, KIT. }\label{fig:cr_candidate}
\end{figure}

\begin{figure}[t!]
\centering
\begin{minipage}{0.54\linewidth}
    \includegraphics[trim={1cm 0cm 4cm 0cm},clip,width=\linewidth]{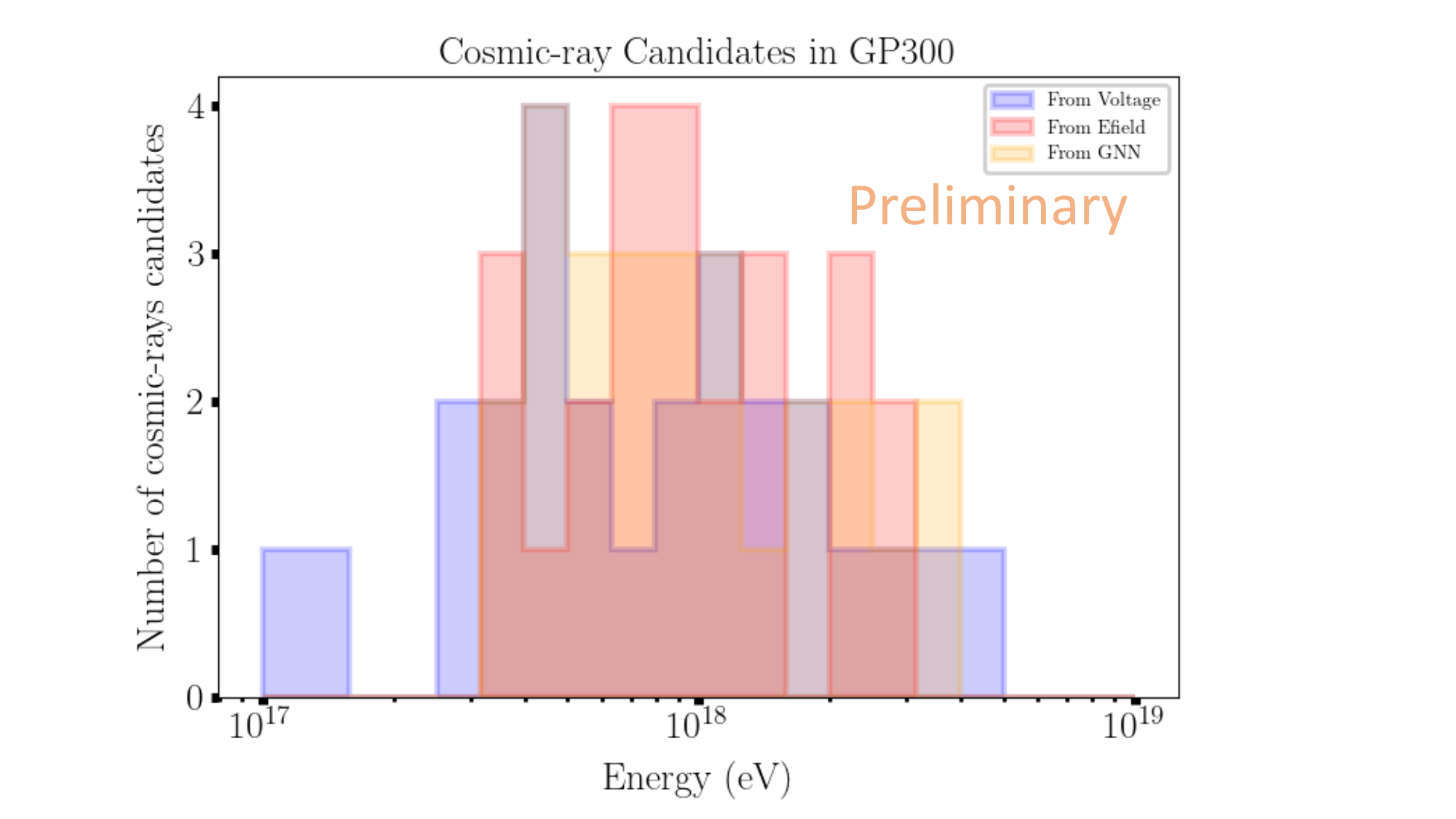}
\end{minipage}%
\hfill
\begin{minipage}{0.45\linewidth}
    \captionof{figure}{Energy spectrum of 26 cosmic-ray candidates computed with 3 independent methods: one fitting the Lateral Distribution Function to the electric field traces\,\cite{ldf}, another fitting the voltage signal with the Amplitude Distribution Function\,\cite{adf_v}, and a third using a Graph Neural Network\,\cite{gnn}. These 26 events involve at least 5 DUs with reconstructed electric field\,\cite{kewen}, correspond to errors on energy lower than $10^{20}$\,eV\ for the LDF fit and yield a $\chi^2$/ndf<25 for the ADF fit. Calibration, candidate selection and energy reconstruction procedures are still being refined. This figure therefore only illustrates the consistency between these 3 methods. }
    \label{fig:spectrum}
\end{minipage}
\end{figure}

The main experimental effort of the GRAND collaboration is taking place in the Gobi Desert, on a large plateau called XiaoDuShan (40.99$^{\circ}$\,N, 93.94$^{\circ}$\,E, 1200\,m a.s.l.), Gansu Province, China. GP300, currently under deployment at this location, will ultimately consist of an array of 289 detection units (DUs) covering a total area of 200\,km$^2$. At present, 65 DUs are deployed—enough to detect a few tens of cosmic rays per day in the energy range of $10^{17}$–$10^{18}$\,eV\cite{sei}. Commissioning of GP300 is nearing completion. During this period, several significant results have been achieved: \\
- a major effort to reduce or mitigate self-generated noise was undertaken in the early months of commissioning. The system has now reached its nominal noise level, with a very homogenous response (see Fig.\,\ref{fig:DUfig}), and GP300 takes full advantage of the excellent electromagnetic conditions at the site. A clear daily modulation of antenna signal noise is observed (see Fig.\,\ref{fig:gal}), corresponding to the transit of the Galactic Plane through the antennas’ field of view. \\
- trigger parameters were optimized, resulting in a measured trigger efficiency of $\sim$75\% for event rates up to 50\,Hz. This was evaluated on a subarray of a few antennas using a beacon generator emitting prompt pulses, similar to those expected from extensive air showers (EAS). Further improvements in trigger efficiency are ongoing. \\
- Wi-Fi data transfer was tested at rates up to ~4\,MB/s for point-to-point communication. With ~20 DUs connected to a single Wi-Fi receiver in the DAQ room, this translates into 200\,kB/s per DU, allowing a maximum second-level trigger rate of 20\,Hz for an event size of 8\,kB. This is twice the nominal rate defined for GP300\,\cite{decoene:hal-02317328}. \\
- the resolution of relative timing is better than 5\,ns, enabling beacon position reconstruction with a precision better than 1\,m and azimuthal angle reconstruction with a resolution around 0.1$^\circ$ for a distant, static ground source~\cite{gp300}. Further refinements in timing and reconstruction are ongoing. \\
- a procedure for cosmic ray identification has been developed and is described in~\cite{cr_search}. While still being optimized, it has already yielded a list of 41 cosmic-ray candidates detected with GP300 between December 2024 and March 2025, for which arrival direction and energy could be reconstructed with three independant methods\,\cite{adf_v,ldf,gnn}. An example is shown in Fig.~\ref{fig:cr_candidate} and the energy spectrum of the 29 candidates with valid energy reconstruction in Fig\,\ref{fig:spectrum}.

\subsection{GRAND@Auger}
The G@A prototype was set up in collaboration with the Pierre Auger Observatory in 2024. It consists of 10 GRANDProto DUs (see Section~\ref{du_design}) deployed on the site of the Auger Engineering Radio Array (AERA), reusing part of its infrastructure—particularly the mechanical supports, power supply, communications, and DAQ room.

The main objective of G@A is to perform autonomous triggering on transient radio events, identify cosmic-ray candidates among them through offline analysis, and search for possible coincidences with Pierre Auger data. The expected statistics are limited by the modest size of the G@A array but this data sample provides a unique opportunity to conduct an event-by-event study of the purity, efficiency, and resolution of EAS reconstructed parameters using GRAND detection.

\begin{figure}[t!]
\centering
\includegraphics[trim={0cm 0 4cm 0cm},clip,width=0.48\linewidth]{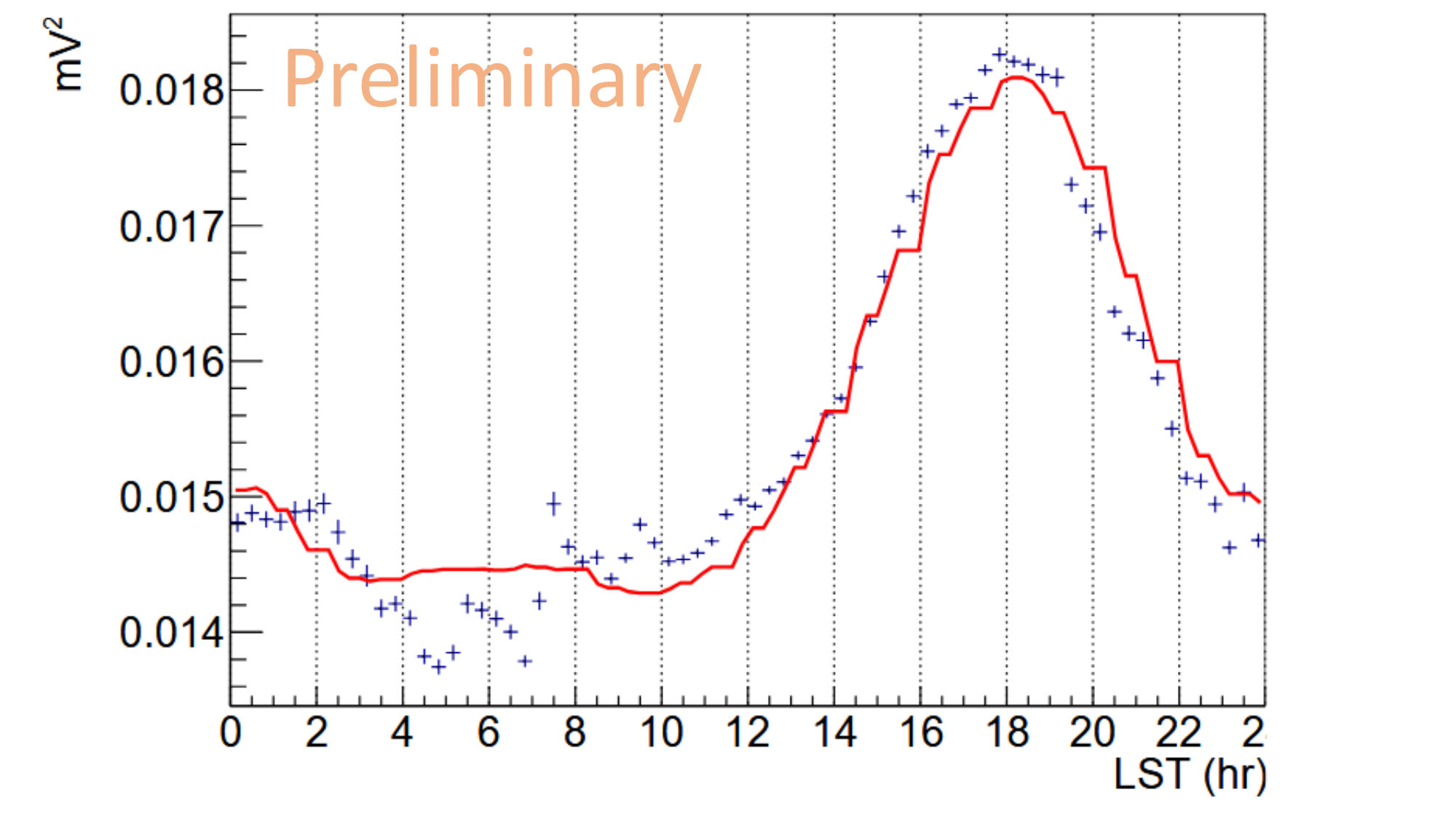}
\includegraphics[trim={2cm 0 2cm 0},clip,width=0.50\linewidth]{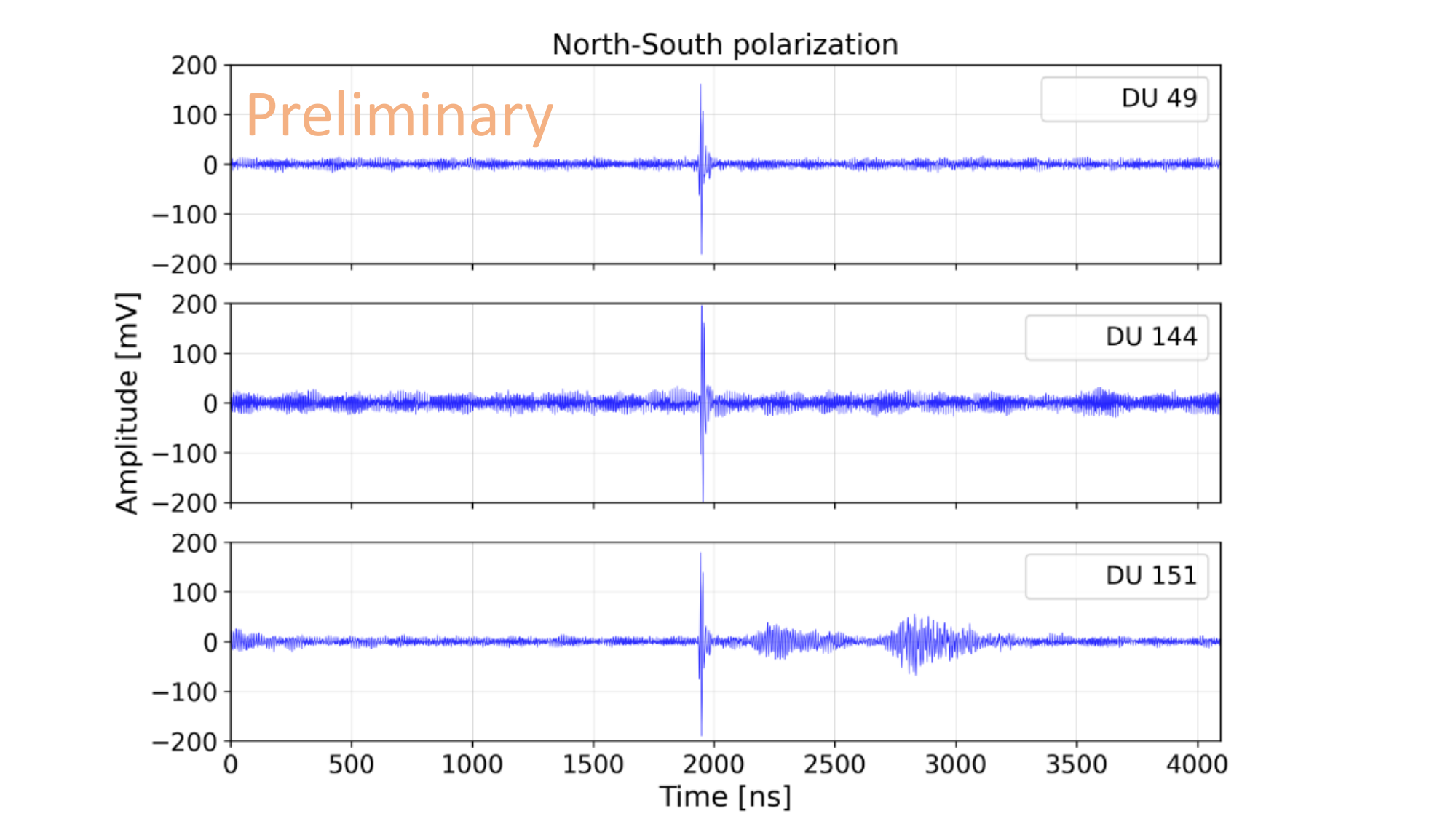}
\caption{Left: standard deviation of the radio signal measured in the 100-200\,MHz band along the East-West arm of DU83 as a function of local sidereal time. The error bars indicate the uncertainty on the mean. The red line represents the galactic simulation fitted to the data, with an additional offset and amplification. Right: the self-triggered event with three G@A detector units observed in coincidence with a cosmic ray detected by Auger. Plots taken from\,\cite{gaa}.  }\label{fig:gaa}
\end{figure}

The GRAND collaboration is still working to optimize the G@A operational parameters. The work carried out so far is reported in~\cite{gaa}: we have observed the Galactic daily modilation  in the 100-200\,MHz band  and G@A has also detected a radio event coincident with a $1.3 \times 10^{19}$\,eV cosmic ray observed by Auger (see Fig.\,\ref{fig:gaa}).

\section{Next steps}
\label{next}
In the short term, the GRAND collaboration will focus on optimizing operational parameters at the GP300 and G@A sites, and on improving the cosmic-ray identification pipeline. The goal is to reach the expected cosmic-ray candidate rates—a few tens per day at GP300 and one per day at G@A—within the next few months. The combined dataset will make it possible to quantify the performance of the GRAND detector in terms of background rejection, EAS detection efficiency, and reconstruction of EAS properties.

Over the next 1-3 years, advanced trigger methods currently under development within the collaboration (see e.g.~\cite{nutrig, denoising}) are expected to be implemented, leading to a lower energy threshold and improved data purity. GP300 will also be completed in this period of time, possibly with design evolutions in preparation for the next phase of the experiment: GRAND10k. GRAND10k will consist of two sub-arrays of $\sim$10,000 DUs, deployed in both hemispheres —most likely in China and Argentina— offering improved sky coverage. It will have potential for neutrino discovery and will also serve to finalize the design of the complete GRAND detector. Various options are being explored to reduce power consumption and optimize detector reliability. The current triggering strategy will also likely be replaced by a distributed architecture. More radical changes are being explored, such as a hybrid design with a phased radio array triggering autonomous antennas~\cite{heron}.

\section{Conclusion}
The GRAND prototypes are coming to life. Our first stand-alone radio detection of cosmic-ray events was achieved during their commissioning phase. With forthcoming optimization of operating conditions and candidate identification, stable operation is expected shortly and shall lead to the validation of the GRAND detection principle. Ongoing developments in detector design, triggering and reconstruction are paving the way toward GRAND10k and, ultimately, the complete GRAND detector: a new era for ultra-high-energy neutrino astronomy.

\bibliographystyle{ICRC}
\setlength{\bibsep}{0pt plus 0.3ex}
{\footnotesize
\bibliography{references}
}
%

\clearpage

\section*{Full Author List: GRAND Collaboration}

\scriptsize
\noindent
J.~Álvarez-Muñiz$^{1}$, R.~Alves Batista$^{2, 3}$, A.~Benoit-Lévy$^{4}$, T.~Bister$^{5, 6}$, M.~Bohacova$^{7}$, M.~Bustamante$^{8}$, W.~Carvalho$^{9}$, Y.~Chen$^{10, 11}$, L.~Cheng$^{12}$, S.~Chiche$^{13}$
% ...
\\
\\
$^{1}$Departamento de Física de Particulas \& Instituto Galego de Física de Altas Enerxías, Universidad de Santiago de Compostela, 15782 Santiago de Compostela, Spain \\
$^{2}$Institut d'Astrophysique de Paris, CNRS  UMR 7095, Sorbonne Université, 98 bis bd Arago 75014, Paris, France \\
$^{3}$Sorbonne Université, Université Paris Diderot, Sorbonne Paris Cité, CNRS, Laboratoire de Physique  Nucléaire et de Hautes Energies (LPNHE), 4 Place Jussieu, F-75252, Paris Cedex 5, France \\
$^{4}$Université Paris-Saclay, CEA, List,  F-91120 Palaiseau, France \\
$^{5}$Institute for Mathematics, Astrophysics and Particle Physics, Radboud Universiteit, Nijmegen, the Netherlands \\
$^{6}$Nikhef, National Institute for Subatomic Physics, Amsterdam, the Netherlands \\
$^{7}$Institute of Physics of the Czech Academy of Sciences, Na Slovance 1999/2, 182 00 Prague 8, Czechia \\
$^{8}$Niels Bohr International Academy, Niels Bohr Institute, University of Copenhagen, 2100 Copenhagen, Denmark \\
$^{9}$Faculty of Physics, University of Warsaw, Pasteura 5, 02-093 Warsaw, Poland \\
$^{10}$Key Laboratory of Dark Matter and Space Astronomy, Purple Mountain Observatory, Chinese Academy of Sciences, 210023 Nanjing, Jiangsu, China \\
$^{11}$School of Astronomy and Space Science, University of Science and Technology of China, 230026 Hefei Anhui, China \\
$^{12}$National Astronomical Observatories, Chinese Academy of Sciences, Beijing 100101, China \\
$^{13}$Inter-University Institute For High Energies (IIHE), Université libre de Bruxelles (ULB), Boulevard du Triomphe 2, 1050 Brussels, Belgium \\
% ...


\subsection*{Acknowledgments}

\noindent
The GRAND Collaboration is grateful to the local government of Dunhuag during site survey and deployment approval, to Tang Yu for his help on-site at the GRANDProto300 site, and to the Pierre Auger Collaboration, in particular, to the staff in Malarg\"ue, for the warm welcome and continuing support.
The GRAND Collaboration acknowledges the support from the following funding agencies and grants.
\textbf{Brazil}: Conselho Nacional de Desenvolvimento Cienti\'ifico e Tecnol\'ogico (CNPq); Funda\c{c}ão de Amparo \`a Pesquisa do Estado de Rio de Janeiro (FAPERJ); Coordena\c{c}ão Aperfei\c{c}oamento de Pessoal de N\'ivel Superior (CAPES).
\textbf{China}: National Natural Science Foundation (grant no.~12273114); NAOC, National SKA Program of China (grant no.~2020SKA0110200); Project for Young Scientists in Basic Research of Chinese Academy of Sciences (no.~YSBR-061); Program for Innovative Talents and Entrepreneurs in Jiangsu, and High-end Foreign Expert Introduction Program in China (no.~G2023061006L); China Scholarship Council (no.~202306010363); and special funding from Purple Mountain Observatory.
\textbf{Denmark}: Villum Fonden (project no.~29388).
\textbf{France}: ``Emergences'' Programme of Sorbonne Universit\'e; France-China Particle Physics Laboratory; Programme National des Hautes Energies of INSU; for IAP---Agence Nationale de la Recherche (``APACHE'' ANR-16-CE31-0001, ``NUTRIG'' ANR-21-CE31-0025, ANR-23-CPJ1-0103-01), CNRS Programme IEA Argentine (``ASTRONU'', 303475), CNRS Programme Blanc MITI (``GRAND'' 2023.1 268448), CNRS Programme AMORCE (``GRAND'' 258540); Fulbright-France Programme; IAP+LPNHE---Programme National des Hautes Energies of CNRS/INSU with INP and IN2P3, co-funded by CEA and CNES; IAP+LPNHE+KIT---NuTRIG project, Agence Nationale de la Recherche (ANR-21-CE31-0025); IAP+VUB: PHC TOURNESOL programme 48705Z. 
\textbf{Germany}: NuTRIG project, Deutsche Forschungsgemeinschaft (DFG, Projektnummer 490843803); Helmholtz—OCPC Postdoc-Program.
\textbf{Poland}: Polish National Agency for Academic Exchange within Polish Returns Program no.~PPN/PPO/2020/1/00024/U/00001,174; National Science Centre Poland for NCN OPUS grant no.~2022/45/B/ST2/0288.
\textbf{USA}: U.S. National Science Foundation under Grant No.~2418730.
Computer simulations were performed using computing resources at the CCIN2P3 Computing Centre (Lyon/Villeurbanne, France), partnership between CNRS/IN2P3 and CEA/DSM/Irfu, and computing resources supported by the Chinese Academy of Sciences.

\end{document}